\documentclass[twocolumn,twocolappendix]{aastex631}

\newcommand{\kpc}{\,\mathrm{kpc}}

\newcommand{\NOMP}{N_{\rm OMP}}
\newcommand{\NMPI}{N_{\rm MPI}}
\newcommand{\Ncore}{N_{\rm core}}

\newcommand{\Msun}{M_{\sun}}

\usepackage[normalem]{ulem}
\usepackage{comment}
\usepackage{threeparttable} 
\usepackage{booktabs}     
\usepackage{amsmath}

\newcommand\edel{\bgroup\markoverwith{\textcolor{red}{\rule[0.5ex]{2pt}{0.8pt}}}\ULon}

\graphicspath{{./figures}}

\shortauthors{Han et al.}

\graphicspath{{./}{figures/}}

\begin{document}

\title{RAMSES-yOMP: Performance Optimizations for the Astrophysical Hydrodynamic Simulation Code RAMSES}

\author[0000-0001-9939-713X]{San Han}
\altaffiliation{Email address: san.han@iap.fr, sanhan@yonsei.ac.kr}
\affiliation{Department of Astronomy and Yonsei University Observatory, Yonsei University, Seoul 03722, Republic of Korea}
\affiliation{Institut d’Astrophysique de Paris, Sorbonne Université, CNRS, UMR 7095, 98 bis bd Arago, 75014 Paris, France}

\author[0000-0003-0225-6387]{Yohan Dubois}
\affiliation{Institut d’Astrophysique de Paris, Sorbonne Université, CNRS, UMR 7095, 98 bis bd Arago, 75014 Paris, France}

\author[0000-0002-6810-1778]{Jaehyun Lee}
\affiliation{Korea Astronomy \& Space Science Institute, Daejeon 305-348, Republic of Korea}

\author[0000-0002-4391-2275]{Juhan Kim}
\affiliation{Korea Institute of Advanced Studies (KIAS) 85 Hoegiro, Dongdaemun-gu, Seoul 02455, Republic of Korea}

\author[0000-0003-2285-0332]{Corentin Cadiou}
\affiliation{Lund Observatory, Division of Astrophysics, Department of Physics, Lund University, Box 43, SE-221 00 Lund, Sweden}
 
\author[0000-0002-4556-2619]{Sukyoung K. Yi}
\altaffiliation{Email address: yi@yonsei.ac.kr}
\affil{Department of Astronomy and Yonsei University Observatory, Yonsei University, Seoul 03722, Republic of Korea}

\begin{abstract}
Developing an efficient code for large, multiscale astrophysical simulations is crucial in preparing the upcoming era of exascale computing.
RAMSES is an astrophysical simulation code that employs parallel processing based on the Message Passing Interface (MPI).
However, it has limitations in computational and memory efficiency when using a large number of CPU cores.
The problem stems from inefficiencies in workload distribution and memory allocation that inevitably occur when a volume is simply decomposed into domains equal to the number of working processors.
We present RAMSES-yOMP, which is a modified version of RAMSES designed to improve parallel scalability.
Major updates include the incorporation of Open Multi-Processing into the MPI parallelization to take advantage of both the shared and distributed memory models.
Utilizing this hybrid parallelism in high-resolution benchmark simulations with full prescriptions for baryonic physics, we achieved a performance increase of a factor of 2 in the total run-time, while using 75\% less memory and 30\% less storage compared to the original code, when using the same number of processors.
These improvements allow us to perform larger or higher-resolution simulations than what was feasible previously.
\end{abstract}

\section{Introduction}\label{sec:intro}
Recent large-scale simulation projects \citep[e.g.,][]{2018MNRAS.473.4077P,2021A&A...651A.109D,2021ApJ...908...11L} have made prominent achievements in the field of astrophysical research, which have been made possible by the advancement of high-performance computing (HPC).
The improvement of HPC facilities has mainly been achieved by increasing the number of processor cores, rather than enhancing the clock speed of a single core.
This growth has necessitated the development of codes optimized for many-core parallelism.

RAMSES \citep{2002A&A...385..337T} is a simulation code designed for solving various astrophysical problems.
The code implements gravitational and hydrodynamical interactions based on the adaptive mesh refinement (AMR) technique that utilizes an Octree system, a structure in which a cubic grid can be recursively subdivided into eight octants to control the spatial resolution of the simulation \citep{1997ApJS..111...73K}.

For large-scale simulations that typically require many computing cores or nodes, RAMSES employs the Message Passing Interface (MPI), a parallel processing library that utilizes the distributed memory model.
In this scheme, the simulation data is distributed to dedicated memory spaces (i.e., MPI domains) on each processor, and when data from different domains are needed, the corresponding data are communicated.
RAMSES relies on a domain decomposition scheme based on the Peano-Hilbert space-filling curve \citep{2000CoPhC.126..330M,2008ApJS..178..179P}, in which three-dimensional domain volumes are mapped onto segments of a one-dimensional space.
This spatial decomposition scheme can reduce data communication overhead, as most of the interactions between spatially adjacent components (particles, cells) in the same domain can be locally processed without requiring inter-domain communication.

If the computational workload is well-balanced across MPI domains, each parallel task can maximize its performance while being remaining completely independent of the others.
A disadvantage of the distributed memory scheme is that it requires good load balancing between MPI domains for the optimal use of processors.
This procedure, called load balancing, is usually achieved through data exchange between domains to reduce the inhomogeneity of the workload among them.

This requires the measurement of computational load across different domains, which is extremely challenging because the amount of workload depends on many factors, such as the number of simulation elements (e.g., grids and particles), the degree of clustering, densities, temperatures, and other physical properties.
Furthermore, the domain remapping process needs to be executed frequently to suppress the development of load inhomogeneity.
Ironically, as the remapping process itself is computationally expensive, a trade-off needs to be made between the overhead caused by remapping and the load imbalance.

The challenge of load balancing becomes even more demanding in high-resolution simulations of multiscale physical processes due to the nonuniform distribution of matter and energy densities across the simulation box.
This results in varying degrees of resolution requirements and, consequently, a heterogeneous distribution of computational workload across the simulation volume.

For example, in a typical galaxy simulation, the central region of the galaxy is more populated by stars than the outer region with a higher stellar velocity dispersion and acceleration.
Accurate calculations of gravity and hydrodynamics in such regions require much higher spatial and temporal resolution and are more computationally intensive compared to the outer regions.
This leads to the implementation of an AMR scheme, which adaptively distributes grids of different sizes depending on the local density conditions.

\begin{figure}[t]
    \begin{center}
        \includegraphics[width=0.47\textwidth]{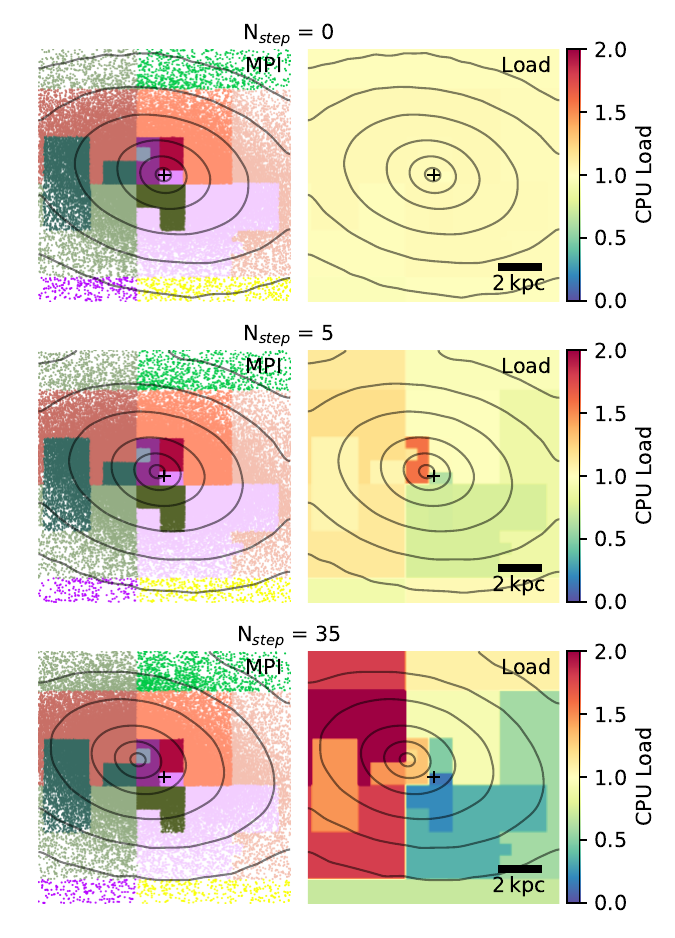}
        \caption{MPI partition map (left column) and load balance map (right column) of a small slice in the region around the central region of a galaxy at multiple time steps in a cosmological hydrodynamic simulation. The black contours represent the stellar density distribution of the galaxy. The black crosses indicate the initial center of the galaxy. The left column shows star particles in an 80-pc-thick slice, color-coded by their associated MPI domain number. The right column shows the CPU load imbalance map in the same slice, which is estimated based on the RAMSES's load cost function for each domain, normalized by the average value. $N_{\rm step}$ represents the number of main steps passed since the last load balancing was conducted.
            \label{fig:load_imbalance}}
    \end{center}
\end{figure}

Fig.~\ref{fig:load_imbalance} shows an example of load balancing with MPI domain decomposition.
The figure shows snapshots of a zoom-in simulation that focuses on a single MW-sized galaxy using RAMSES, with spatial resolution of $68\,\rm pc$ at $z\sim0$.
Each row represents multiple snapshots of a thin slice passing through the central region of the galaxy, where the volume is divided into segments along the Peano-Hilbert space-filling curve.
$N_{\rm step}$ (indicated in the subtitle) represents the number of coarse time steps elapsed since the last domain decomposition.
The left panels show the particles color-coded based on their respective MPI domains.
The right panels show the estimated computational load in each MPI domain represented by colors, normalized by the mean load.
The black contours represent the density distribution of the galaxy, which gradually shifts towards the upper left over time.
The cross symbol marks the initial position of the galaxy center.

At the time of the domain decomposition ($N_{\rm step}=0$), a significant number of MPI domains are allocated to the central region to share the high computational load, resulting in an equal distribution of CPU loads across domains.
However, as the galaxy moves, the distribution of particles among domains changes, increasing the load imbalance with time ($N_{\rm step}=5$ and $35$).
This load imbalance increases waiting time for processors for the MPI synchronization step, which lead to increase in total execution time.
To relieve this problem, more frequent load balancing is required.
However, frequent load balancing yields a serious computational overhead, and the simulation consequently suffers a loss of performance.

Employing a large number of MPI domains introduces additional issues.
Firstly, as the number of MPI domains increases, the areas of the boundary interfaces between neighboring domains also increases.
This essentially leads to an increase in the total amount of data that needs to be communicated between domains, which can result in network bottlenecks during boundary interactions.
Secondly, even with optimal workload balancing among domains, there may still be some variability in the memory usage across domains.
These issues are especially prominent for high-resolution hydrodynamical simulations, which are usually accompanied by more crowded regions with matter and particles demanding higher computational accuracies and shorter time steps.
This leads to large number of domains to be assigned in high-resolution regions to accommodate their workloads, while few MPI domains dominate the low-resolution regions with very small computational workloads.
This inhomogeneity often limits the maximum size of simulations and the minimum number of MPI domains due to the upper bound of memory space available to each domain.
The state-of-the-art galaxy formation models also tend to add extra information in their simulations, such as metallicities, chemical species, and dusts.
These additions result in an increase in overall memory usage, and tighten the memory constraints of the run.

One possible approach to relieve these issues is to implement Open Multi-Processing (OMP) directives.
Since OMP is orthogonal to MPI in the parallel processing space, we can easily scale the number of CPU cores in a two-dimensional parallel framework.
When the total number of CPU cores is fixed, we have one degree of freedom to adjust either the number of MPI domains or the number of OMP threads to maximize the efficiency of the parallel performance.

Another advantage of OMP is the flexibility it offers in distributing computational resources.
The frequency of domain decomposition needs to be determined carefully because the procedure itself introduces additional computational overhead.
In RAMSES, the period is usually set to a certain number of time steps, which is sufficient for the simulation to develop workload inhomogeneity due to the evolving distribution of matter, making the parallel execution slower.
In contrast, the OMP threads in a domain are allocated at the point when the master thread executes the calculation.
This allows for a more efficient allocation of cores by accounting for the dynamically changing conditions of the simulation.

This paper is organized as follows.
In Section~\ref{sec:data}, we describe modifications made to enhance parallel efficiency of the RAMSES code.
Section~\ref{sec:results} presents the results of benchmark simulations to test the computational and memory performance of the new code compared to the original code.
We also perform a consistency test to verify the reliability of results from the new code.
Finally, in Section~\ref{sec:conclusion}, we present our conclusions.

\section{Methods}\label{sec:data}
\subsection{the RAMSES code}
We use a state-of-the-art version of RAMSES for our modifications.
The code shares prescriptions similar to those of the NewHorizon simulation~\citep{2021A&A...651A.109D}, including star formation based on a thermo-turbulent model \citep{2017MNRAS.466.4826K}, mechanical supernova feedback \citep{2014ApJ...788..121K}, evolution and feedback of supermassive black holes \citep{2012MNRAS.420.2662D,2021A&A...651A.109D}.
It also includes several updated physical processes compared to earlier simulations, such as stellar winds, Type Ia supernovae, the evolution of chemical species, and the formation and destruction of dust species \citep{2024A&A...687A.240D}.
The implemented physics will be described in more detail by Han et al. (2024, in prep), where we employ the new code for large-scale zoom-in simulations of a galaxy cluster.
The code also includes a tracer particle scheme based on a Monte Carlo method \citep{2019A&A...621A..96C}.
From now on, we will refer to the new code as RAMSES-yOMP, or simply yOMP.

\subsection{OMP implementation}
We extensively employed OMP for optimizing the time-intensive subroutines of the code.
In most cases, these subroutines involve do-loops across every grid or particle, making them directly applicable to OMP.
We applied fork-join design, where the master thread opens the sub-threads at the initiation of each loop and collects their result to shared memory at the end of the loop.

\begin{table*}[t]
\centering
\begin{threeparttable}
\caption{
List of subroutines that are responsible for the majority of the execution time in each category.
\label{tab:subroutines}}
\begin{tabular}{lll}
\toprule
Category & Major subroutine(s) & Role \\
\midrule
\hline
Hydro & \texttt{godunov\_fine} & Hydrodynamics \\
 & \texttt{cooling\_fine} & Gas cooling \\
\hline
Gravity & \texttt{phi\_fine\_cg} & High-resolution gravity \\
 & \texttt{multigrid\_fine} & Low-resolution gravity \\
\hline
Particles & \texttt{rho\_fine} & Particle density \\
 & \texttt{synchro\_fine \& move\_fine} & Particle movement \\
 & \texttt{star\_formation} & Star formation\\
 & \texttt{mechanical\_feedback\_fine} & SNe Type II \\
 & \texttt{mechanical\_feedback\_fine\_snIa} & SNe Type Ia \\
 & \texttt{stellar\_winds} & Stellar winds \\
 & \texttt{create\_sink} & Black hole creation \\
 & \texttt{grow\_bondi} & Black hole accretion \\
 & \texttt{AGN\_feedback} & Black hole feedback \\
\hline
Load balance & \texttt{load\_balance} & Load balancing \\
 & \texttt{refine\_fine} & AMR refinement \\
\hline
Misc. & \texttt{newdt\_fine} & Time stepping \\
\bottomrule
\end{tabular}
\end{threeparttable}
\end{table*}

For the efficient operation of multiple threads that are assigned to each MPI domain, OMP must be applied to as many subroutines as possible.
Therefore, the implementation is applied to most of the time-intensive regions of RAMSES, including both on-grid and sub-grid subroutines.
The majority of subroutines where OMP is implemented are listed in Table~\ref{tab:subroutines}, categorized by their characteristics.

OMP directives have two different scheduling options, \texttt{static} and \texttt{dynamic}.
The \texttt{static} option forces the number of tasks distributed per thread to be even.
We use this option in subroutines where the task operates on a grid-wise basis, and the computational cost is similar between loops.
The \texttt{static} option assigns each thread a fixed chunk of data that is spatially adjacent in the simulation space, which can benefit from various optimizations such as automatic vectorization and caching.
The \texttt{dynamic} option assigns tasks to threads whenever a thread completes its current task, at the cost of a small overhead due to the task scheduling operations.
This option is applied when the computational cost between loops varies significantly.
In RAMSES, this corresponds to subroutines involving particles, where the loops are executed based on a grid, but the actual computation is performed on particles attached as a linked list to the parent grid at that location.

\subsection{Additional improvements}\label{sec:imp}
In addition to implementing OMP, yOMP includes several key improvements that significantly contribute to the performance enhancement.
The original RAMSES utilizes the conjugate gradient (CG) method for solving the Poisson equation, which is required for the computation of the small-scale gravitational potential at the cell level.
The solver starts with an initial guess to solve the Poisson equation and calculates its errors.
It then iteratively updates the guess until the convergence conditions for errors are satisfied.
For each iteration, a global MPI communication is required, which represents the largest performance bottleneck of the algorithm.

We replaced the native CG solver with a preconditioned pipelined conjugate gradient (PPCG) solver \citep[Algorithm 4]{GHYSELS2014224}, which incorporates preconditioning and pipelined iteration method.
We use a preconditioner matrix from \citet[Equation 24]{5452414}, generalized to a three-dimensional incomplete stencil consisting of seven nonzero points, with the central point set to $13/12$ and six neighboring points each set to $1/6$.
Compared to the standard CG solver, the PPCG solver significantly reduces the total number of iterations needed for the convergence of the solutions, thereby decreasing the number of the global MPI synchronization steps and alleviating resulting communication bottlenecks.

Another modification was made to the cost function for measuring the computational workload in domain decomposition.
In the RAMSES simulation, the load balancing subroutine aims to flatten the distribution of predetermined cost weights across different domains.
The cost in each domain is evaluated as
\begin{equation}\label{eq:cost}
    C=\sum^{\ell_{\rm max}}_{\ell=\ell_{\rm min}}w(\ell)(kN_{{\rm grid}}(\ell) + N_{{\rm part}}(\ell)),
\end{equation}
where $\ell_{\rm min}$ and $\ell_{\rm max}$ indicate the minimum and maximum AMR levels of the simulation, respectively.
$N_{\rm grid}(\ell)$ and $N_{\rm part}(\ell)$ represent the number of grids and particles at level $\ell$, respectively.
Each time the grid level increases by 1, the grid size is halved.
The particle level is determined to be equal to the level of the parent grid where the particle is located.
$w(\ell)$ is the weighting factor that applies to each level.
In a cost-weighted load balancing scheme, the weighting factor is generally defined as the number of fine time steps that need to be taken at the level $\ell$ to complete one coarse time step, e.g., $w(\ell+1)=2 w(\ell)$ for sub-cycled time steps ($\Delta t(\ell+1)=\Delta t(\ell)/2$), but $w(\ell+1)=w(\ell)$ for single time steps ($\Delta t(\ell+1)=\Delta t(\ell)$).
$k$ is a weight parameter that determines the relative cost between grids and particles. By default, its value is set to 10.
However, we have found that this value may not be optimal for simulations with state-of-the-art baryonic physics, which implement more complex phenomena related to particles such as Type Ia and II supernovae, stellar winds, black hole feedback, etc.
This also indicates that the ideal value of $k$ needs to be chosen adaptively, depending on the astrophysical prescription of the simulation.
For the yOMP version, we chose a value of $k=2.5$, corresponding to 4 times higher weights for particles.

\section{Results}\label{sec:results}
\subsection{Reduction of Execution Time}\label{sec:improvement}
To evaluate the parallel efficiency of the code, we performed benchmark simulations with cosmological initial conditions that have been evolved down to $z\sim4$, in a cube with a comoving volume of $(12.5\, h^{-1}\,\rm Mpc)^3$ paved with a minimum resolution of $14 \,\rm kpc$ ($\ell_{\rm min}=8$), with the best spatial resolution of $54\,\rm pc$ ($\ell_{\rm max}=16$).
In simulations employing hybrid parallelism that utilize both MPI and OMP, the total number of working cores, denoted as $\Ncore$, corresponds to the product of the number of MPI domains, denoted as $\NMPI$, and the number of OMP threads denoted as $\NOMP$, i.e.,
\begin{equation}\label{eq:ncore}
    \Ncore = \NMPI \times \NOMP.
\end{equation}
For a fair comparison between different configurations of hybrid parallelism, we fixed $\Ncore$ and varied $\NMPI$ and $\NOMP$.
The test was conducted on a multi-node system with Intel Xeon Phi 7250 processors, with 68 cores per node, based on the Knights Landing architecture, interconnected via a 100Gbps Omni-Path Architecture network.

\begin{figure}[t]
    \begin{center}
        \includegraphics[width=0.47\textwidth]{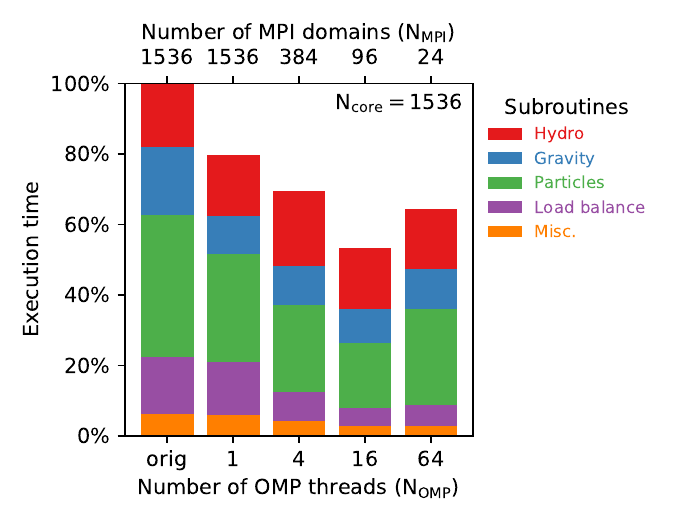}
        \caption{
        Parallel performance of yOMP with different numbers of OMP threads and the original code, in terms of the scaled wall clock time.
        The number of cores used for parallel jobs has been fixed across all runs ($\Ncore=1536$).
        Each bar represents a simulation with different combinations of OMP and MPI numbers.
        Execution times are normalized to the original code.
        Each color corresponds to a set of categorized subroutines responsible for calculating specific astrophysical processes in RAMSES (see Table~\ref{tab:subroutines} for detail).
        The optimal performance is achieved with 16 threads, reducing the execution time to $\sim53\%$ of the original code.
        In general, subroutines related to gravity and particles exhibit significant improvements, while the speedup in refinement and load balancing is due to the reduced number of MPI domains.
        \label{fig:perf_improvements_24}}
    \end{center}
\end{figure}

Fig.~\ref{fig:perf_improvements_24} represents the relative execution time of yOMP compared to the original RAMSES code.
The first bar (labeled as ``orig'') represents the result from the original code, while the following bars show the results from yOMP with different $\NOMP=(1,4,16,64)$, and accordingly $\NMPI=(1536,384,96,24)$ for a fixed number of total working cores $\Ncore=1536$.
The result with $\NOMP=1$ indicates the MPI-only run with the yOMP code.
This test is conducted to assess the performance gains from the additional modifications described in Sec. \ref{sec:imp} that are not related to hybrid parallelism.
We have grouped the subroutines in RAMSES into several categories based on their specific tasks in the simulation. A summary of these subroutines by category is provided in Table~\ref{tab:subroutines}.

The yOMP code exhibits the best performance with 16 threads, achieving a minimum execution time of $\sim53\%$ of that of the original code.
This corresponds to a speed boost of about a factor of 2.
Performance deteriorates when using 64 threads, indicating that, with a fixed number of cores, there is a ``sweet spot'' for selecting $\NMPI$ and $\NOMP$ to achieve optimal performance.
In this performance test, we observe diminishing marginal performance gains with increasing $\NMPI$ and $\NOMP$.
Hybrid parallelism helps alleviate this issue by utilizing lower values for both $\NMPI$ and $\NOMP$, thus preserving the linear scalability of the parallel structure.
Although the optimal values may vary depending on simulation size, it is expected that there is an optimal combination of $\NMPI$ and $\NOMP$ that yields the best performance.

\subsection{Performance Boost by Categories}\label{sec:category}
\begin{figure}[t]
    \begin{center}
        \includegraphics[width=0.47\textwidth]{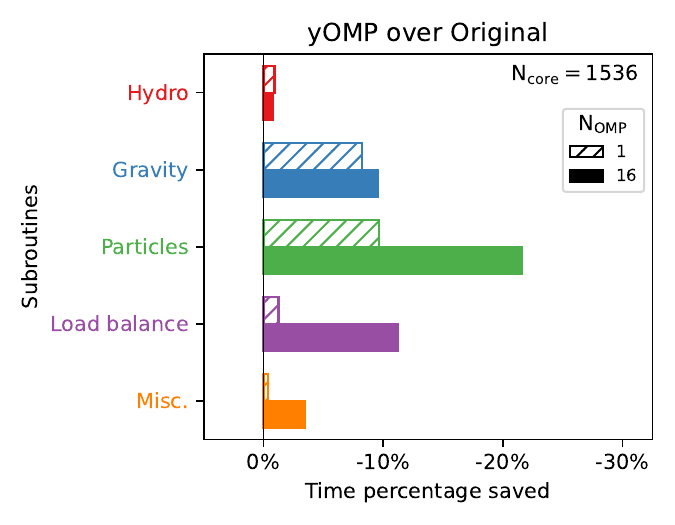}
        \caption{
        Performance improvements measured from Fig.~\ref{fig:perf_improvements_24} across different
        categories of subroutines.
        The x-axis represents the reduced fraction of total execution time compared to the original code byyOMP and using different OMP threads, 1 (hatched bars), 16 (filled bars).
        Color schemes represent different categories of subroutines specified in Table~\ref{tab:subroutines}.
        \label{fig:perf_by_subroutine_24}}
    \end{center}
\end{figure}

Fig.~\ref{fig:perf_by_subroutine_24} shows the performance gain obtained with yOMP for different categories of subroutines, where the same color scheme is adopted as in Fig.~\ref{fig:perf_improvements_24}.
Bars with hatches represent the time savings achieved by using yOMP in an MPI-only mode ($\NOMP=1$), and filled bars represent the combination of $\NOMP=16$ and $\NMPI=384$, which showed the best performance in Fig.~\ref{fig:perf_improvements_24}.

The best time saving is found in the ``Particles'' category, which includes calculations related to simulation particles.
This means that the particles have quantities that are well parallelized in nature.
Nearly half of the time savings is due to the poor performance of the cost function in the original code.
The relative cost between particles ($k$ in Equation~\ref{eq:cost}) and cells has been chosen to be 10 for the original RAMSES simulation, meaning that the cost weight of one cell is regarded as that of 10 particles in the process of load balancing.
We find this results in the simulation overestimating the cost of cells relative to particles, so we reduc the ratio to 2.5.

The other half of the performance improvement appears to come from increasing OMP threads, well demonstrating the efficiency of hybrid parallelism.
As indicated in Fig.~\ref{fig:load_imbalance}, the MPI-only domain decomposition has limitations in achieving ideal load balancing, especially for particles whose highly-inhomogeneous distributions are hard to catch in the current Peano-Hilbert domain decomposition scheme.
The improvement is likely due to better allocation of computational resources across simulation volumes, resulting from implementing a shared memory configuration.

The ``Load balancing'' category that directly deals with the domain decomposition process accounts for the second-greatest time savings.
The improvement is mainly led by a smaller number of MPI domains, which simplifies domain decomposition and reduces the total amount of communications.
This is the natural result of increasing $\NOMP$, which reduces $\NMPI$ with a fixed number of $\Ncore$ according to Equation~\ref{eq:ncore}.

\begin{figure}[t]
    \begin{center}
        \includegraphics[width=0.47\textwidth]{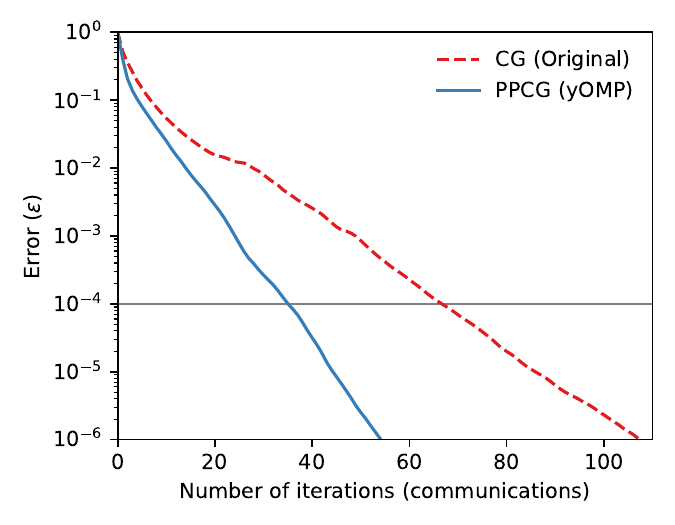}
        \caption{
        The relative error $\varepsilon$ of the Poisson equation solution as a function of the number of iterations, comparing two different algorithms, CG (red dashed line) and PPCG (blue solid line).
        The gray horizontal line corresponds to the conventional level of accuracy that is required to solve the equation ($10^{-4}$).
        The new PPCG method decreases the number of iterations needed to achieve a specific level of accuracy by a factor of $\sim2$.
        \label{fig:iter_poisson}}
    \end{center}
\end{figure}

The ``Gravity'' category consists of subroutines that solve Poisson's equation in a given density distribution.
Most of the performance improvement comes from changing the algorithm from CG to PPCG.
An example of a comparison between two algorithms in solving the Poisson equation is presented in Fig.~\ref{fig:iter_poisson}.
The y-axis in the figure indicates the error of the solution to the Poisson equation, which is evaluated as
\begin{equation}\label{eq:eps}
\varepsilon_i = \sqrt{(r_i \cdot u_i) / (r_0 \cdot u_0)},
\end{equation}
where $r$ is the residual of the linear system $r_i = b - Ax_i$, $u$ is the residual after applying the preconditioner, i.e., $u_i = M^{-1} r_i$, $r_0$ and $u_0$ are residuals of the initial guess $x_0$.
For the case of CG algorithm, $M^{-1}$ can be regarded as an identity matrix.
The x-axis shows the number of iterations applied in the algorithm.
The horizontal line indicates the accuracy that is required for the solution of the equation.
Compared to the native CG algorithm (red dashed line) of the original code, the new PPCG algorithm (blue dashed line) significantly reduces the number of iterations by a factor of $\sim1.86$ to reach a certain relative error of the solution ($\varepsilon=10^{-4}$).
Reducing the number of iterations also reduces the amount of communication that occurs during each gravity-solving step, which is one of the main bottlenecks of the algorithm, resulting in a significant performance improvement.

The ``Hydro'' category shows negligible change in the execution time, suggesting that the parallel scalability for the hydro solver is similar for MPI and OMP, regardless of the choice of $\NMPI$ and $\NOMP$, at least for our benchmark simulation.
This also indicates that the hydrodynamics solver, which operates on each cell in the simulation, may not be significantly affected by load imbalance.
As shown in Equation~\ref{eq:ncore}, an increase in OMP threads comes at the expense of the number of MPI tasks. The results of Fig.~\ref{fig:perf_by_subroutine_24} for ``Hydro'' category can be interpreted in this light, with MPI-based parallelization being as efficient as our proposed hybrid strategy.

To evaluate the adaptability of the code, the same benchmark test has been conducted on an operating system with a different CPU architecture and interconnection network. 
The result is presented in Appendix~\ref{sec:icelake}.

\subsection{Savings in Memory and Disk Space Utilization}

The consumption of random access memory (RAM) and disk space in simulations has been a major obstacle to the fast calculation of hydrodynamic simulations.
This issue stems from the inhomogeneous distribution of memory in different domains, illustrated in Fig.~\ref{fig:load_imbalance}.
In the figure, the MPI domains show different sizes, with smaller sizes located in the denser regions, while the domains in the surrounding regions are larger in size.
This is because more resources are assigned to more computationally intensive regions with smaller resolution elements requiring a larger number of time steps to reach the solution, resulting in relatively smaller domains and less data possession.
This results in great inequality between domains in terms of their private data usage, and most of the low-level particles and cells, which have very small load weights, are monopolized by a few domains in the coarse regions.
At the startup, RAMSES pre-allocates an equal amount of memory for particles and grids in each domain, ensuring that the allocation can accommodate the domain with the maximum memory utilization.
Simulations with a larger $\NMPI$ tend to be more vulnerable to the inhomogeneities that arise from particle clustering or deep Octree cells in each domain, which may lead to inefficient memory usage.

There are also disadvantages to using a large $\NMPI$ in the efficiency of disk space utilization.
When RAMSES saves grid data as a snapshot, it includes not only the main grid owned by each domain but also grids from adjacent domains in the boundary region, known as ``reception grids.''
As the number of MPI domains increases, the number of these redundant reception grids also increases, resulting in higher storage requirements for the snapshot savings.

\begin{figure}[t]
    \begin{center}
        \includegraphics[width=0.47\textwidth]{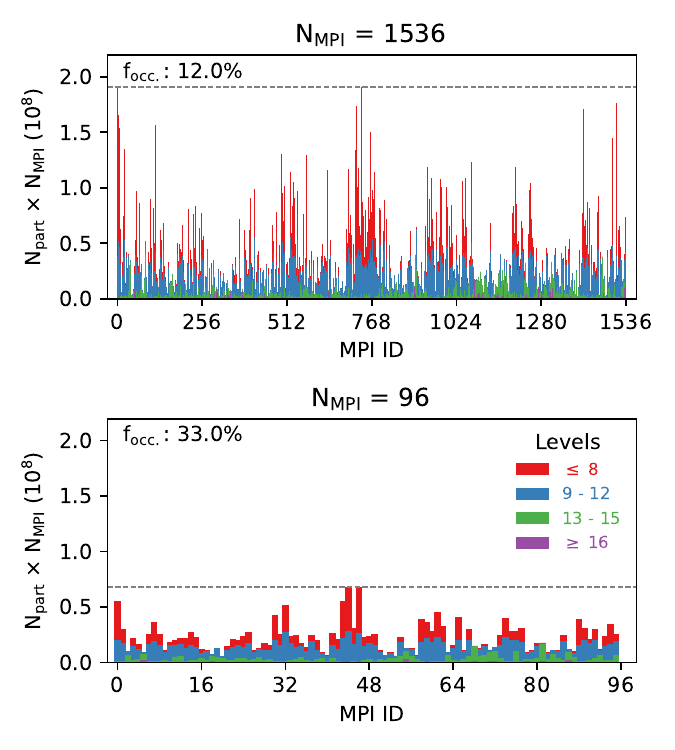}
        \caption{Distribution of particles in the benchmark simulation with two different numbers of MPI domains ($\NMPI$).
            The x-axis shows the MPI rank ID for domains.
            The height of each bar indicates the number of particles owned in each domain, multiplied by the number of MPI domains, for the fair comparison of memory usage between simulations with different numbers of domains.
            The color indicates the particles with different levels.
            The horizontal dashed line represents the largest number of particles in a domain, which corresponds to the minimum allocation size required for the particle data to accommodate the simulation data.
            The mean filling percentage ($f_{\rm occ}$) using the minimum required allocation is shown in the upper left side of each panel as an indicator of memory utilization efficiency.
            The bottom panel with $\NMPI=96$ requires a much smaller allocation space, meaning that memory can be utilized more efficiently in this configuration.
            \label{fig:cpumap_vertical}}
    \end{center}
\end{figure}

Hybrid parallelism can resolve these issues by reducing the number of MPI domains, which can in turn save RAM and disk space usage for the simulation.
Fig.~\ref{fig:cpumap_vertical} visualizes the distribution of memory usage for particles in different MPI configurations.
The two panels show the number of particles in each domain with different numbers of MPI domains: $\NMPI=1536$ (upper panel) and $\NMPI=96$ (lower panel).
The heights of the bars show the number of particles owned by each domain, multiplied by $\NMPI$ to enable a fair comparison of memory usage between the upper and lower panels.
The colors indicate the particle level, where particles with higher levels are more load-intensive and therefore have greater weights in load balancing.
The horizontal dashed line indicates the minimum required memory space to be allocated to accommodate the particles in all domains.
The occupation fraction of particles, $f_{\rm occ}$, which is the fraction of memory that has been used compared to the minimum memory required for the simulation to accommodate the MPI domain with the largest number of particles, is shown for the case where the minimum required allocation is applied to the simulation.
The occupation fraction is 33.0\% for the $\NMPI=96$ case and 12.0\% for $\NMPI=1536$, indicating that simulations with a smaller number of domains use memory more efficiently.
Memory savings have also been achieved with the distribution of grid numbers across different MPI domains in a similar manner.

\begin{figure}[t]
    \begin{center}
        \includegraphics[width=0.47\textwidth]{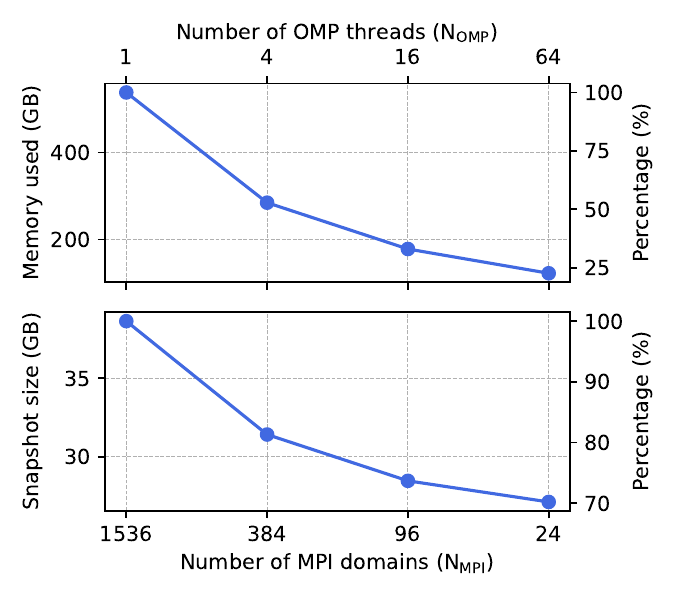}
        \caption{RAM and disk space usage by a different number of OMP threads and MPI domains, with the number of cores fixed ($\Ncore=1,536$).
            More threads lead to fewer domains, which reduces memory imbalance and redundancy.
            Up to $\sim77\%$ of RAM and $\sim30\%$ of disk space savings have been achieved as a result.
            \label{fig:mem_mpi}}
    \end{center}
\end{figure}

Fig.~\ref{fig:mem_mpi} shows the RAM (Panel a) and disk space usage (Panel b) of the simulation with different $\NMPI$.
We compared the configurations that use hybrid parallelization to the original code using $\NMPI=1536$.
When using $\NMPI=24$, RAM usage is reduced to $\sim23\%$, which corresponds to a $\sim4.3\times$ reduction.
Also, disk space usage is reduced to $\sim70\%$, which corresponds to a $\sim1.4\times$ reduction.
When using $\NMPI=96$, which has shown the best computational performance in Fig.~\ref{fig:perf_improvements_24} RAM usage is reduced to $\sim30\%$  which corresponds to a $\sim 3.3\times$ reduction.
The disk space usage is also reduced to $\sim75\%$, which corresponds to a $\sim 1.3\times$ reduction.

\subsection{Consistency Test}

\begin{figure*}[t]
    \begin{center}
        \includegraphics[width=1.0\textwidth]{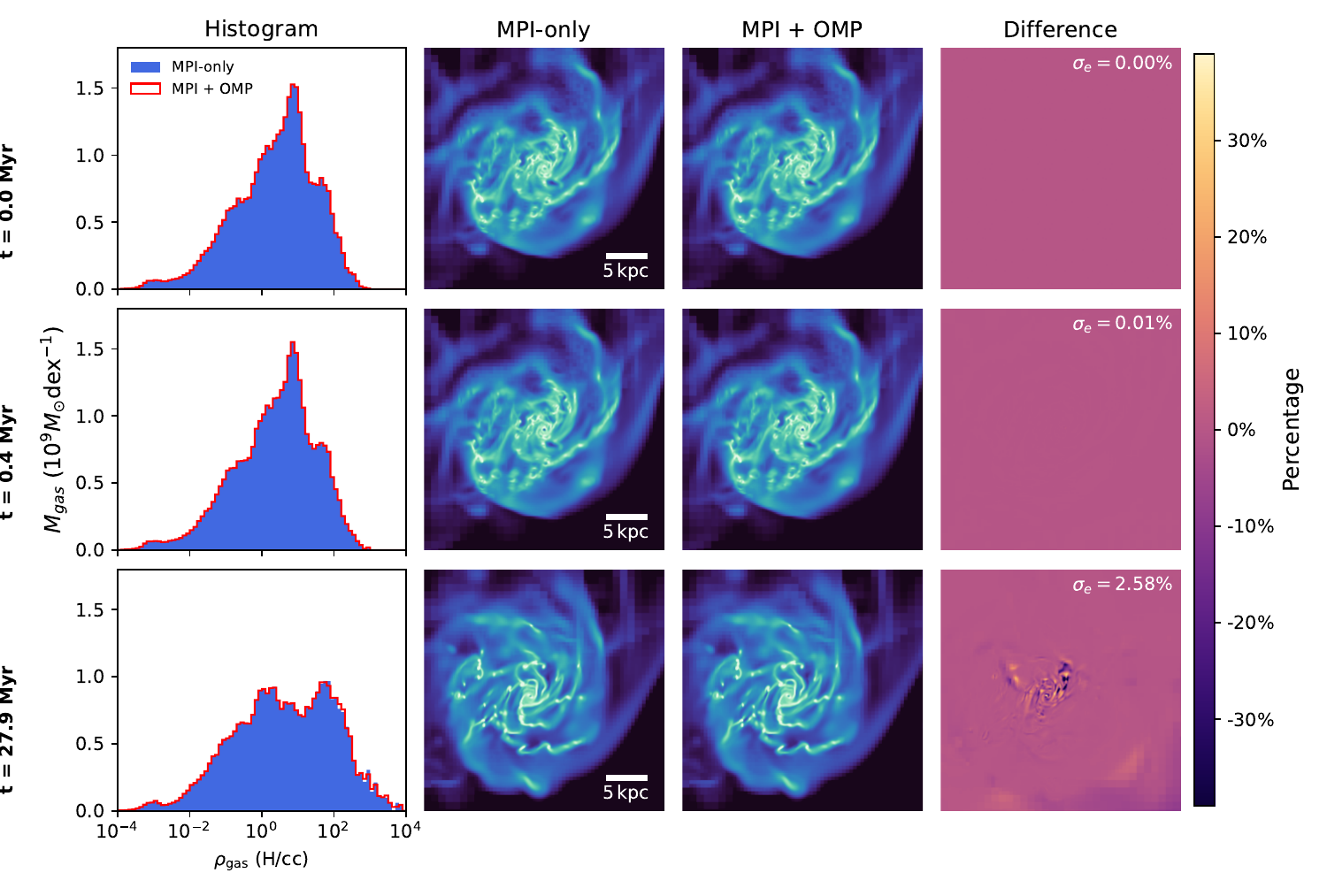}
            \caption{Comparison between MPI-only and hybrid parallelism (MPI + OMP) runs of a zoom-in simulation of a spiral galaxy, starting from the same initial state.
            Each row represents snapshots at different times since the initial state. The elapsed time ($t$) is shown in units of Myr on the left side of each row.
            The first column shows the comparison of gas density histograms in the same volume.
            The second and third columns (MPI-only, MPI + OMP) show the evolution of the gas column density distribution for the two different configurations.
            The fourth column (Difference) indicates the fractional difference in the gas density map between the two configurations, with a side length of $30\,\kpc$.
            The standard deviation of the difference ($\sigma_{\rm diff}$) is shown in the upper right corner of each difference map.
            \label{consistency_test}}
    \end{center}
\end{figure*}

The OMP implementation in this work has been focused to reproduce the behavior of the simulation in MPI-only mode. To ensure the reliability of the code, it is important to verify whether the simulation using the new scheme reproduces reasonably similar results to those obtained from the MPI-only mode.
A consistency test has been performed with the simulation of a single galaxy starting from $z=1.5$, which has been generated from a zoom-in cosmological run with baryonic physics activated.
The target galaxy has a stellar mass of $10^{10.8}\,\Msun$ at the start of the simulation.
We ran simulations with two configurations: MPI-only mode ($\NMPI=48$) and hybrid mode ($\NMPI=48$ and $\NOMP=4$), starting from the same initial state of the galaxy.
During the consistency test, we had to halt the star formation process because its stochastic nature made it challenging to reproduce consistently in the OMP run.

Fig.~\ref{consistency_test} presents the results of the consistency test.
Each row shows different snapshots of the simulation, with the first row representing the initial state.
The elapsed time since the start is shown at the left end of each row ($t$).
The first column shows the comparison of mass-weighted gas cell density distributions for two different configurations.
The second and third columns display the gas column density distribution of MPI-only and MPI + OMP (hybrid) runs, respectively.
The fourth column (Difference) shows the contrast map between the MPI-only and MPI + OMP configurations and the standard deviations of the difference ($\sigma_{\rm diff}$).

The two simulations start from an identical initial condition.
In the second row, which represents a snapshot after one coarse step has passed since the start of the run, the gas map still appears largely the same, but there is a slight deviation.
In the third row, the apparent structure of the gas map looks very similar for both configurations.
However, the difference map shows a clear deviation between MPI-only and hybrid runs.

It seems the results of the two configurations slowly diverge from each other, showing a small deviation in the central regions of the galaxy.
However, it is not surprising that hybrid parallelization produces results that differ from the MPI-only run.
These differences are mainly caused by the change in the order of arithmetic operations during the OMP-specific optimizations, which may accumulate different patterns of round-off errors with subsequent time evolution.
This round-off error usually happens with various levels of compiler optimizations, but this kind of error is beyond what is typically obtained due to stochastic behavior of galaxy simulations \citep{2008MNRAS.387..397T,2019ApJ...871...21G,2019MNRAS.482.2244K}.

We further discuss the effect of round-off error in Appendix~\ref{sec:nvec}, where we present density and contrast maps of MPI-only runs with different orders of arithmetic operations.

\section{Conclusion}\label{sec:conclusion}
In this paper, we introduced RAMSES-yOMP, a new modified version of the astrophysical hydrodynamic simulation code RAMSES.
RAMSES-yOMP implements a hybrid parallelization scheme, adding OMP on top of the preexisting MPI structure.
The new code also includes two additional performance improvements: a preconditioned pipelined conjugate gradient method for solving the Poisson equation, and an enhanced load balancing scheme for particles.
The major results of this paper can be summarized as follows:

\begin{itemize}
\item In the benchmark simulation utilizing $1536$ cores, the new code has shown a reduction in execution time to $\sim53\%$, which corresponds to approximately a factor of $2$ speedup compared to the original code.
This reduction arises from multiple factors, including better load balancing between cores, less overhead for domain decomposition, and the improved Poisson equation solver.
\item Memory usage has been reduced up to $\sim23\%$ (a factor of $\sim4$), and disk space usage up to $\sim70\%$ (a factor of $\sim1.4$).
Using the number of OMP threads that has shown the best performance ($\NOMP=16$), memory and disk space are reduced to $\sim33\%$ and $\sim74\%$, respectively.
Memory savings are achieved by reducing the number of MPI domains, which levels out the variability of memory usage across different domains.
The reduction in disk space usage arises from the reduced redundancy of saved grid data in the MPI structure.
\item The new code with hybrid parallelization reasonably reproduces the results of the MPI-only run, although the results slightly diverge over time due to a different pattern of round-off errors in arithmetic operations.
\end{itemize}

RAMSES-yOMP shows great improvements in terms of parallel performance and memory efficiency.
The hybrid parallelism implemented by yOMP provides various advantages.
The new code offers the potential for performing large-scale simulations that effectively utilize more than $\sim10 000$ cores.
This adds a great opportunity for the astrophysical community that demands increasingly challenging amount of computational resources.
The reduction in MPI domains allows for the efficient use of memory, enabling simulation to be performed in facilities with small RAM sizes without the need for intentionally idling cores to secure memory space.
The secured memory space can also be utilized for the allocation of information that can be useful for storing additional physical properties, such as chemical species, dust, and radiation within cell data (Han et al. 2025, in prep.).
This combination of performance improvements allows for performing large-scale simulations that were previously not feasible due to parallel inefficiencies of the code.

Nevertheless, several potential bottlenecks with the yOMP code remain in the hydrodynamic simulations with baryonic physics, mainly arising from the parallel implementation of particle-based subroutines that interact with the grid structure.
The most prominent sectors include the density estimation using the cloud-in-cell (CIC) algorithm and mechanical supernova feedback \citep{2014ApJ...788..121K}.
These subroutines share the characteristic that each thread can simultaneously access and update overlapping grid locations.
To prevent possible data corruption from the race conditions, the code must include atomic, critical, or lock operations, which introduce a slowdown in the simulation.

The severity of this access crowding problem depends on the update frequency of overlapping shared arrays by different threads, which scales with the number of threads and is likely the primary cause of performance deterioration in the case of $\NOMP=64$ as shown in Fig.~\ref{fig:perf_improvements_24}, resulting in the presence of a ``sweet spot''.
This issue can be alleviated by allocating a temporary array for each thread and performing most computations in private memory before applying summarized updates to shared memory all at once with reduced frequency.
This strategy has been employed in most of the OMP implementations in the code.
However, some subroutines involve complex iterative features that do not permit parallel updates to the main memory properties.
This is the case for mechanical supernova feedback, which involves an algorithm that infers and updates the target and neighboring grids in a single sequential execution, thereby enforcing single-threaded operations.
A full OMP implementation of this algorithm may require relinquishing the constraint of consistently reproducing single-threaded behavior.

The flexibility of OMP allows the number of cores to be changed during a simulation run, which was not possible in the original version of RAMSES.
This ability adds significant convenience, especially for simulation projects that utilize a large amount of computing resources, where users need to adapt to resource availability.
The published version of the code is available at \cite{han_2024_14013854}.
The live version of the code is available at a git repository\footnote{\hyperlink{https://github.com/sanhancluster/RAMSES-yOMP}{https://github.com/sanhancluster/RAMSES-yOMP}}.

\section*{Acknowledgements}
S.K.Y. acknowledges support from the Korean National Research Foundation (NRF-2020R1A2C3003769 and NRF-2022R1A6A1A03053472). 
This work was granted access to the HPC resources of KISTI under the allocations KSC-2022-CRE-0344 and KSC-2023-CRE-0343.
The large data transfer was supported by KREONET which is managed and operated by KISTI. 


\bibliography{references}{}
\bibliographystyle{aasjournal}



\restartappendixnumbering
\appendix
\section{Results from different operating system}\label{sec:icelake}
In the main text of this paper, we employed a multicore system based on the deprecated Knights Landing architecture, which is not directly comparable to the CPUs used in the latest HPC systems.
The parallel performance of the code may depend on the specification of the operating system, such as single core performance and quality of the interconnection network.
For this reason, we repeated the benchmark test described in sections \ref{sec:improvement} and \ref{sec:category} on a different system that uses Intel Xeon Platinum 8360Y processors, with 72 cores per node, based on the Ice Lake architecture and interconnected via a Mellanox QM8790 InfiniBand network.
The CPUs in this system offer better per-core performance compared to the Knights Landing architecture used in the main benchmark test.

\begin{figure}[h]
    \begin{center}
        \includegraphics[width=0.47\textwidth]{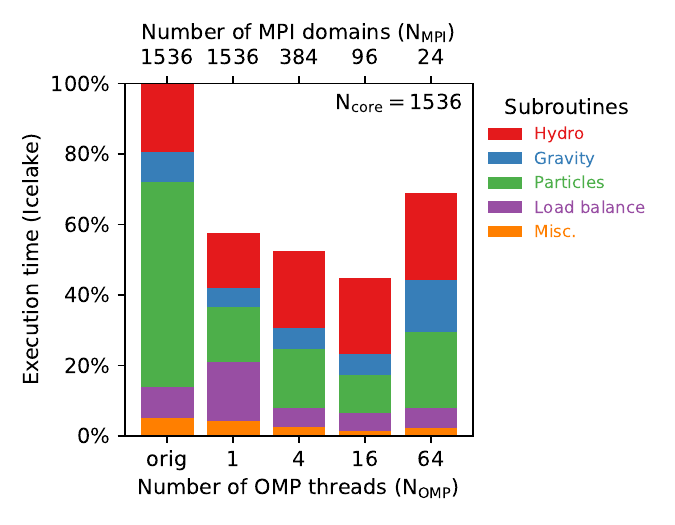}
        \caption{
        Equivalent to Fig.~\ref{fig:perf_improvements_24}, but with using Ice Lake processors. The amount of execution time taken for the benchmark simulation, using original code (orig) and yOMP code with different number of threads.
        Colors indicate the different categories of subroutines (see Table \ref{tab:subroutines}).
        The best performance has been achieved with 16 OMP threads.
        \label{fig:perf_improvements_24_olaf}}
    \end{center}
\end{figure}
\begin{figure}[t]
    \begin{center}
        \includegraphics[width=0.47\textwidth]
        {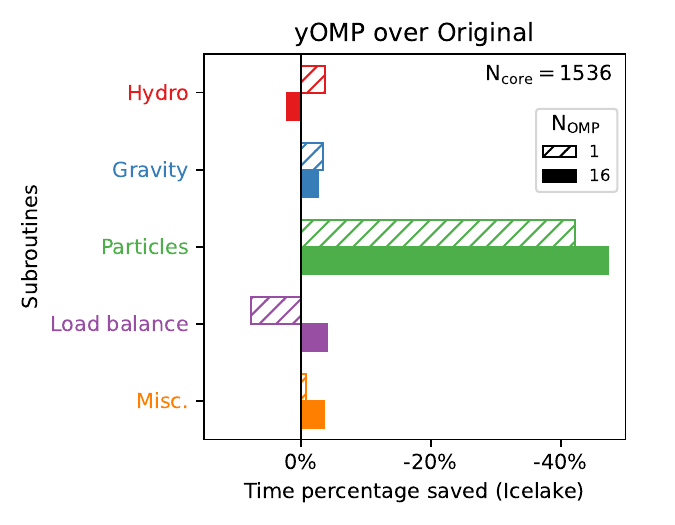}
        \caption{
        Equivalent to Fig.~\ref{fig:perf_by_subroutine_24}, but with using different (Ice Lake) processors.
        The x-axis indicates the fraction of time percentage reduced over original by using the yOMP code.
        Colors indicate different categories of subroutines (see Table \ref{tab:subroutines}).
        Bars with hatches indicate the yOMP run with MPI-only mode, while the filled bars indicate the result using 16 OMP threads.
        \label{fig:perf_by_subroutine_24_olaf}}
    \end{center}
\end{figure}

Fig.~\ref{fig:perf_improvements_24_olaf} and Fig.~\ref{fig:perf_by_subroutine_24_olaf} are equivalent to Fig.~\ref{fig:perf_improvements_24} and Fig.~\ref{fig:perf_by_subroutine_24}, respectively, both compares the percentage of execution time relative to the original code, in which former is comparison in different number of threads and latter is the comparison of subroutines assigned to different categories in Table \ref{tab:subroutines}.

Compared to the previous benchmark run, the new architecture yields a similar reduction in execution time.
Peak performance is observed at $\NOMP=16$, reducing the total execution time to approximately $\sim45\%$ of the original code.
The largest speedup is achieved in the ``Particle'' category, which significantly reduces execution time from the original to $\NOMP=1$ mode. This indicates that the speedup stems from modifying the weighting factor $k$, which enables a more efficient distribution of particle load across MPI processors.
Conversely, the ``Load balance'' category shows a slight increase in execution time when employing yOMP, likely due to the increased exchange of information resulting from a more concentrated distribution of MPI domains caused by particle-oriented load balancing.
Overall, the tests indicate that yOMP consistently outperforms the original code, regardless of the operating system.

\section{Effect of Round-off Errors}\label{sec:nvec}
\begin{figure*}[t]
    \begin{center}
        \includegraphics[width=0.9\textwidth]{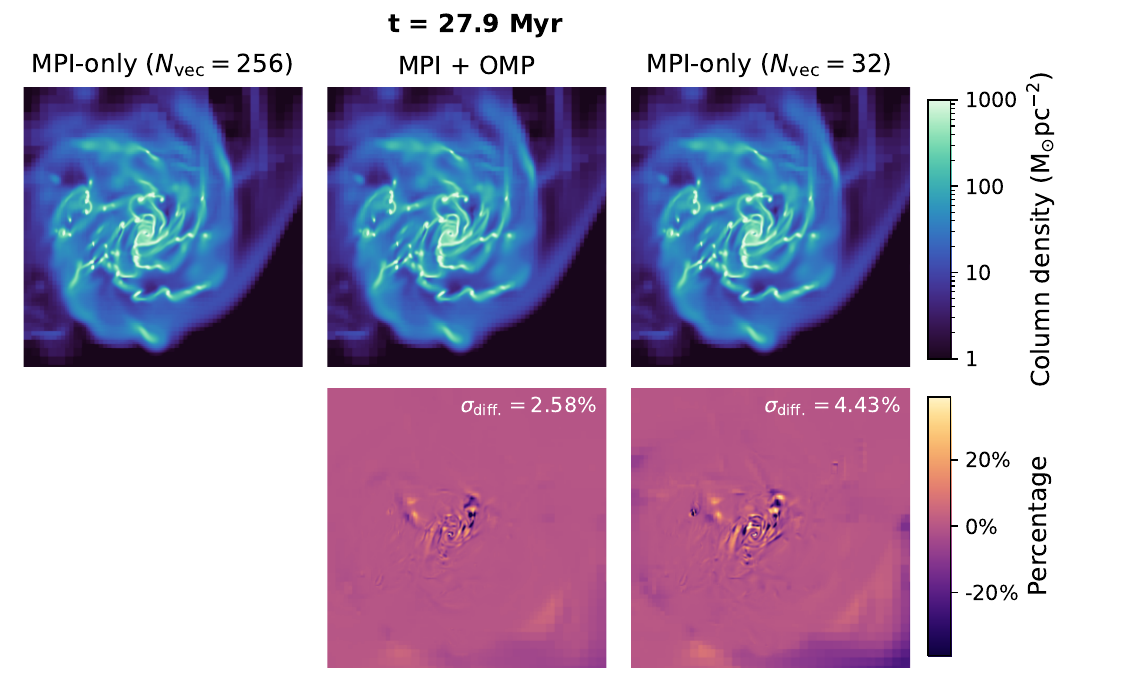}
        \caption{The comparison of density map (first row) and contrast map (second row) of different configurations compared to the base MPI-only run (first column) in the zoom-in galaxy simulation, same as used in Fig.~\ref{consistency_test}.
        The first and second columns show density and contrast maps of MPI-only and MPI + OMP run, same as presented in Fig.~\ref{consistency_test}.
        The third column shows an MPI-only run, but with different vectorization size ($N_{\rm vec}$) applied.
        The contrast map and standard deviation of the difference ($\sigma_{\rm diff}$) show MPI-only run with different $N_{\rm vec}$ have comparable deviation to MPI + OMP run.
        The MPI + OMP run shows less deviation compared to the MPI-only with different $N_{\rm vec}$.
        \label{consistency_test_nvec}}
    \end{center}
\end{figure*}
Fig.~\ref{consistency_test_nvec} shows the consistency test, similar to Fig.~\ref{consistency_test}, but with one additional MPI-only configuration that uses different number of vectorization size ($N_{\rm vec}$).
$N_{\rm vec}$ determines the general number of simulation components that are passed to the subroutines.
In principle, this does not affect the analytic algorithm of the calculation.
However, choosing a different $N_{\rm vec}$ value may change the order of arithmetic operations and result in different pattern of round-off errors to be produced.
We used the $N_{\rm vec}=256$ for the base MPI run in Fig.~\ref{consistency_test} and $N_{\rm vec}=32$ for another MPI-only run in the third column to intentionally generate round-off error.
The resulting contrast map shows a similar difference pattern to the MPI + OMP run but with larger deviations.
This indicates that the deviation between the results from MPI-only and MPI + OMP runs is well within the variation range produced from stochastic selection of round-off errors.

\end{document}